\documentclass[12pt]{article}

\usepackage{cite}
\usepackage{epsf}
\usepackage{graphicx}
\usepackage{psfrag}
\usepackage[dvips]{epsfig}

\usepackage{verbatim}

\usepackage[cp866]{inputenc}
\usepackage[english]{babel}
\usepackage{amssymb,indentfirst}

\makeatletter
\renewcommand \thesection {\@arabic\c@section.}
\renewcommand\thesubsection   {\thesection\@arabic\c@subsection.}
\renewcommand\thesubsubsection{\thesubsection\@arabic\c@subsubsection.}
\makeatother

\def\lineup#1{\mbox{$\raise1.0ex\hbox{--} \kern-0.6em#1$}}

\def\starup#1{\mbox{$\raise1.6ex\hbox{$*$} \kern-0.5em#1$}}
\def\starupp#1{\mbox{$\raise1.8ex\hbox{$*$} \kern-1.0em#1$}}
\def\staruppp#1{\mbox{$\raise1.0ex\hbox{$*$} \kern-0.5em#1$}}
\def\sstarup#1{\mbox{\scriptsize $\raise1.8ex\hbox{$*$} \kern-.7em#1$}}

\def\ttildeup#1{\mbox{$\raise0.0ex\hbox{\Large $\; \tilde{}$} \kern-0.45em#1$}}
\def\bbar#1{\mbox{$\raise-0.4ex\hbox{\Large $\; \bar{}$} \kern-0.35em#1$}}

\begin{document}

\title{
Vector leptoquark mass limits and \\ 
branching ratios of $ K_L^0, B^0, B_s \to l^+_i l^-_j  $ decays with \\ 
account of fermion mixing in leptoquark currents
 }

\author{ A.~D.~Smirnov\thanks{E-mail: asmirnov@univ.uniyar.ac.ru}\\
{\small\it Division of Theoretical Physics, Department of Physics,}\\
{\small\it Yaroslavl State University, Sovietskaya 14,}\\
{\small\it 150000 Yaroslavl, Russia.}}
\date{}
\maketitle

\begin{abstract}
\noindent                             
The contributions of the vector leptoquarks of Pati-Salam type to the branching 
ratios of $K_L^0, B^0, B_s \to l \, l^{\prime}$ decays are calculated with account 
of the fermion mixing in the leptoguark currents of the general type. 
Using the general parametrizations of the mixing matrices  
the lower vector leptoquark mass limit $ m_V > 86 \,\, TeV $ is found   
from the current experimental data on these decays.    
The branching ratios of the decays $B^0, B_s \to l \, l^{\prime}$  
predicted at $m_V = 86 \,\, TeV$ are calculated.  
These branching ratios for the decays $ B^0, B_s \to \mu^+ \mu^-, e \mu $ 
are close to the experimental data whereas those for the decays  
$B^0 \to e^+ e^-, e \tau, \mu \tau$, and  $B_s \to e^+ e^-$ 
are by order of $2\div4$ less than their current experimental limits.
For the decays $B_s \to  e \tau, \mu \tau$ these branching ratios 
are of order $10^{-10}$ and $10^{-9}$ respectively.  
The predicted branching ratios will be usefull in the current and 
future experimental searches for these decays. 

\vspace{5mm}
\noindent
\textit{Keywords:} Beyond the SM; four-color symmetry; Pati--Salam; leptoquarks; 
B physics; rare decays.   

\noindent
\textit{PACS number:} 12.60.-i

\end{abstract}




The search for a new physics beyond the Standard Model (SM) is one
of the aims of the current experiments at the LHC.  
There is a lot of new physics scenarios (such as supersymmetry, 
left-right symmetry, two Higgs model, extended dimension models, etc.) 
which are now under experimental searches at the LHC.  

One of the possible variants of new physics can be induced by the well known 
Pati-Salam idea on the possible four color symmetry between 
quarks and leptons of the vectorlike type regarding leptons as the quarks 
of the fourth color in frame 
of the gauge group $G_{PS} = SU_V(4)\times SU_L(2)\times SU_R(2)$ \cite{PS}.
This group has three gauge coupling constants related to the strong electromagnetic 
and weak coupling constants and can be regarded as 
an intermediate stage in 
the symmetry breaking of the GUT group $SO(10)$~\cite{FM,CEG,GN} 
embedded in the more large group $E_6$~\cite{GRS,BN}.  
The four color symmetry of the vectorlike type immediately 
predicts new gauge particles - 
the vector leptoquarks $V_{\alpha}, \alpha=1,2,3$ which belong to the multiplet  
$(15, 1, 1)$ of the group $G_{PS}$ and form the color triplet $(3, 1)_{2/3}$  
of the SM group $G_{SM} = SU_c(3)~\times~SU_L(2)~\times~U(1)$.     
 
In general case the leptoquarks as the vector or scalar particles carrying 
both the  baryon and lepton numbers appear in many models and can led 
to varied new physics effects, the comprehensive review of physics effects 
generated by leptoquarks can be found in~\cite{DFGKK}.   
In the last years the vector $(3, 1)_{2/3}$ leptoquarks are used to explain
 the known anomalies in the semileptonic $B$ meson decays with simultaneous 
satisfaying the other experimental data. Using the conventional Pati-Salam 
vector leptoquarks this goal is achieved by the appropriate choice of the vector leptoquark 
couplings to fermions~\cite{SMG}, by account of the fermion mixing in leptoquark currents 
of the special form~\cite{AFG}, in frame of new gauge leptoquark model extending 
the  $SU(4)$ four color group by the additional factor $SU(3)^{\prime}$~\cite{LGN}, 
by extending the fermion sector of the Pati-Salam model~\cite{CCL}, 
in a three-site gauge model using the Pati-Salam gauge group 
for each fermion generation~\cite{BCFI}.   
Another approach to the explanations of the $B$-decay anomalies is attempted 
in the model with the composite leptoquarks which is based 
on the Pati-Salam SU(4) group 
in the context of a new strongly interacting sector~\cite{BMS,BT}. 
In this approach the vector leptoquarks couple primarily to the fermions 
of the third generation with their couplings to the the fermions of 
the first two generations beeng suppressed.

The lower mass limits for the vector leptoquarks from their direct searches are 
of about or less $1 \,\, TeV$. The essentially more stringent lower mass limits 
are resulting from the rare decays of pseudoscalar mesons. The most stringent of them 
are resulting from the $K^0_L \to e^{\mp} \mu^{\pm}$ decay and with neglect 
of fermion mixing in leptoquark currents are of order of $2\,000 \,\, TeV$   
\cite{VW, KM1,KM2,AD-Mpla2007,AD-YaF2008}.

It should be noted however that the fermion mixing in leptoquark currents 
is quite natural.  
It is as natural as the fermion mixings in the week currents which are described by 
the well known matrices $V_{CKM}$ and $U_{PMNS}$ in the quark and   
lepton sectors respectively.         
Indeed, the mass eigenstates of left- and right-handed quarks and leptons  
$Q^{L,R}_{pa\alpha}$, $l^{L,R}_{ia}$ can enter to interactions with gauge 
and scalar fields in general case through the superpositions 
\begin{eqnarray}
{Q'}^{L,R}_{pa\alpha}=\sum_{q}(A^{L,R}_{Q_a})_{pq} \, Q^{L,R}_{qa\alpha} ,
\,\,\,\,\,\,\,
{l'}^{L,R}_{ia}=\sum_{j}(A^{L,R}_{l_a})_{ij} \, l^{L,R}_{ja}, \; \label{eq:fmix}
\end{eqnarray}
where  $A^{L,R}_{Q_a}$ and $A^{L,R}_{l_a}$ are unitary matrices describing 
the fermion mixing and diagonalizing the mass matrices of quarks and leptons, 
$p, q, i, j = 1,2,3, $... are the quark and lepton generation idexes, 
$ a = 1, 2 $ and $ \alpha = 1, 2, 3 $ are the $ SU_{L}(2) $  and
$ SU_{c}(3) $ indexes,
$Q_{q1} \equiv u_q=(u,c,t)$, $Q_{q2} \equiv d_q=(d,s,b)$ are
the up and down quarks, $l_{j1} \equiv \nu_{j}$
are the mass eigenstates of neutrinos and
$l_{j2} \equiv l_{j}=(e^{-}, \mu^{-}, \tau^{-})$
are the charged leptons.
In the weak interaction the matrices $A^{L,R}_{Q_a}$ and $A^{L,R}_{l_a}$ form 
the CKM and PMNS matrices as 
$C_Q = (A^L_{Q_1})^+ A^L_{Q_2}=V_{CKM}, \,\,\,\,  
C_\ell = (A^L_{\ell_1})^+  A^L_{\ell_2} = U_{PMNS}^+$. 
In analogous way in the interaction of quarks and leptons with leptoquarks 
the matrices $A^{L,R}_{Q_a}$ and $A^{L,R}_{l_a}$ led to specific matrices        
\begin{eqnarray}
 K^{L,R}_a=(A^{L,R}_{Q_a})^+A^{L,R}_{\ell_a}  
\label{eq:KLRa}
\end{eqnarray} 
of the fermion mixing in leptoquark currents.  
The fermion mixing in leptoquark currents can essenially lower the mass limits 
on leptoquark masses. The current experimental data on the decays 
\begin{eqnarray}
K_L^0, B^0, B_s \to l^+_i l^-_j 
\label{eq:KL0B0Bsdecays}
\end{eqnarray} 
give now the possibility to obtain new lower mass limits for the leptoquarks 
with account of the fermion mixing in leptoquark currents.     

In this paper the contributions of the vector leptoquarks of Pati-Salam type 
to the decays~(\ref{eq:KL0B0Bsdecays}) are calculated and analysed with account 
of the fermion mixing in leptoquark currents of the general form and 
the corresponding new lower mass limit for the vector leptoquarks is 
obtained from the current data on these decays. With account of this lower 
mass limit the expected branching ratios of these decays are calculated 
and discussed.   

The interaction of the vector leptoquarks with down fermions which 
are responsible for the decays (\ref{eq:KL0B0Bsdecays}) can be written 
in general case as  
\begin{eqnarray}
  \emph{L}_{Vdl} &=& \frac{g_4}{\sqrt{2}} \big\lbrace (\bar{d}_{p \alpha} [(K^{L}_2)_{pi} \gamma^{\mu}P_L +  
(K^{R}_2)_{pi} \gamma^{\mu}P_R ] l_i) V_{\alpha \mu} + h.c. \big\rbrace ,
\label{eq:lagrVdl}
\end{eqnarray}
where  $g_4 = g_{st}(M_c)$ is the $SU_V(4)$ gauge coupling constant related 
to the strong coupling constant at the mass scale $M_c$ of the $SU_V(4)$ symmetry 
breaking, $P_{L,R}=(1\pm \gamma_5)/2$ are the left and right operators of fermions  
and $K^{L,R}_2$ are the mixing matrices (\ref{eq:KLRa}) for down fermions. 
It should be noted that in the case of the chiral ($K^L_2 \ne K^R_2$) mixing  
the interaction (\ref{eq:lagrVdl}) of vector leptoquarks 
with quarks and leptons is not purely vectorlike. 

Denoting the sums of the branching ratios of charge conjugated final states as  
\begin{eqnarray}
&& 
Br_V(P \to e^+ \mu^-)+ Br_V(P \to \mu^+ e^-) \equiv Br_V(P \to e \mu ),   
\label{eq:Br_VPemu}  
\\
&& 
Br_V(P \to e^+ \tau^-)+ Br_V(P \to \tau^+ e^-) \equiv Br_V(P \to e \tau ) ,  
\\
\label{eq:Br_VPetau}  
&& 
Br_V(P \to \mu^+ \tau^-)+ Br_V(P \to \tau^+ \mu^-) \equiv Br_V(P \to \mu \tau )   
\label{eq:Br_VPmutau}  
\end{eqnarray}
the branching ratios $Br_V(P \to l \, l^{\prime})$
of the decays of pseudoscalar mesons $P =( K^0_L, \, B^0, \, B_s )$ 
into lepton-antilepton pairs  
$l \, l^{\prime} = e^+ e^-, \,\,\, \mu^+ \mu^-, \,\,\,  e \mu, \,\,\,  
e \tau, \,\,\, \mu \tau, \,\,\, \tau^+\tau^- $  
induced by the vector leptoquarks $V$  
in the case of neglecting the electron and muon masses 
\begin{eqnarray}
&& \hspace{-15mm} 
m_e, m_{\mu} \ll m_{\tau},  {m}_{K^0},  {m}_{B^0}, {m}_{B_s}
\label{eq:neglectmemmu}  
\end{eqnarray}
can be written as     
\begin{eqnarray}
&& \hspace{-15mm} Br_V(P \to l \, l^{\prime}) =  
B_P \, \beta_{P,ll^{\prime}}^2 \hspace{5mm} 
\mbox{for} 
\hspace{3mm} l \, l^{\prime} = e^+ e^-, \, \mu^+ \mu^-, \, e \mu, 
\label{eq:Br_VPll1f}  
\\
&& \hspace{-15mm} Br_V(P \to l \, \tau) =  
B_P \, (1-m_{\tau}^2/m_P^2)^2 \, \beta_{P,l \tau}^2 \hspace{5mm}     
\mbox{for} 
\hspace{3mm} l \,\tau = e \tau, \,\,\, \mu \tau,   
\label{eq:Br_VPltauf}  
\\
&& \hspace{-15mm} Br_V(P \to \tau^+\tau^-) =  
B_P \, \sqrt{1-4 m_{\tau}^2/m_P^2  } \,\,
\big [\,\beta_{P,\tau^+\tau^-}^2-(m_{\tau}^2/m_P^2) \, |k^L_{P,33}-k^R_{P,33}|^2 \,\big ],
\label{eq:Br_VPtauptaumf}  
\end{eqnarray}
where
\begin{eqnarray}
&& \hspace{-15mm} B_P = 
\frac{m_{P} \, \pi \,\alpha^2_{st}(M_c) \, f_{P}^2 \, \bbar{m}_{P}^2 \, 
(R_{P}^V)^2 }{2 \, m_V^4 \, \Gamma_P^{tot}  }
\label{eq:BP}  
\end{eqnarray}
are the typical branching ratios of these decays 
and $\beta_{P,ll^{\prime}}^2$, $|k^L_{P,33}-k^R_{P,33}|^2$ are the mixing factors 
depending on the mixing matrices $K^{L}_2$, $K^{R}_2$.     
The enterring into (\ref{eq:BP}) form factors~$f_{P}$ 
parametrize the matrix elements of the axial and pseudoscalar quark currents 
in the standard way, the factors $R_{P}^V$ 
account the gluonic corrections to the pseudoscalar quark current,   
$ \bbar{m}_{P}= m_{P}^2/(m_{d_p}+m_{d_q}) $, $m_{P}$,  $\Gamma_P^{tot}$ are 
the mass and total width of $P$ meson and $m_{d_p}, m_{d_q}$ are the masses 
of its valency quarks, $m_V$ is the mass of the vector leptoquark.    

With account of the definitions (\ref{eq:Br_VPemu})--(\ref{eq:Br_VPmutau}) 
the mixing factors $\beta_{P,ll^{\prime}}^2$ have the form  
\begin{eqnarray}
&& 
\hspace{00mm}  
\beta_{P,e^+ e^-}^2 = \beta_{P,11}^2 , \hspace{13mm} 
\beta_{P,\mu^+ \mu^-}^2 = \beta_{P,22}^2 , \hspace{12mm}
\beta_{P,\tau^+\tau^-}^2 = \beta_{P,33}^2, \hspace{8mm} 
\label{eq:beta2Pll1a}  
\\
&& 
\beta_{P,e \mu}^2 = \beta_{P,12}^2 + \beta_{P,21}^2 , \hspace{8mm} 
\hspace{-4mm}  \beta_{P,e \tau}^2 = \beta_{P,13}^2  + \beta_{P,31}^2, \hspace{3mm}      
\beta_{P,\mu \tau}^2 = \beta_{P,23}^2  + \beta_{P,32}^2, \hspace{3mm} 
\label{eq:beta2Pll1b}  
\end{eqnarray}
where the mixing factors $\beta_{P,i j}^2$ are related to the matrix elements 
$(K^{L,R}_2)_{pi}, \, (K^{L,R}_2)_{qj}$ 
of the mixing matrices $K^{L}_2$, $K^{R}_2$.

The mixing matrices $K^{L}_2$, $K^{R}_2$ are the unitary $3\times3$ matrices 
and in general case each of them can be parametrized by three angles and six phases as 
\begin{eqnarray}
K^{L,R}_2 \hspace{-1mm} = e^{i\varphi^{L,R}_0} \hspace{-1mm}  \left( \begin{array}{ccc}
 c_{12}c_{13}e^{i\varphi_1} & s_{12}c_{13}e^{i\varphi_2} & s_{13}e^{i\varphi_3}   \\
\hspace{-3mm}   (-s_{12}c_{23}-c_{12}s_{23}s_{13}e^{i\delta})e^{i\varphi_{21}}   & 
\hspace{-3mm}  (c_{12}c_{23}-s_{12}s_{23}s_{13}e^{i\delta})e^{i\varphi_{22}} & 
\hspace{-3mm}  s_{23}c_{13}e^{i\varphi_{23}}  \\
\hspace{-3mm}   (s_{12}s_{23}-c_{12}c_{23}s_{13}e^{i\delta})e^{i\varphi_{31}} & 
\hspace{-3mm}  (-c_{12}s_{23}-s_{12}c_{23}s_{13}e^{i\delta})e^{i\varphi_{32}}  & 
\hspace{-3mm}  c_{23}c_{13}e^{i\varphi_{33}} 
\end{array} \hspace{-2mm}  \right)^{\hspace{-2mm} L,R}, 
\label{eq:KLR2}
\end{eqnarray}
where
\begin{eqnarray}
&& 
\hspace{-15mm} \varphi^{L,R}_{21} = ( \varphi_1+\varepsilon)^{L,R} , \hspace{5mm}   
\varphi^{L,R}_{22} =  (\varphi_2+\varepsilon)^{L,R},  \hspace{5mm} 
\varphi^{L,R}_{23} =  (\varphi_3+\delta+\varepsilon)^{L,R},    
\label{eq:phi2i}  
\\
&& 
\hspace{-15mm} \varphi^{L,R}_{31} = 
 -(  \varphi_2 \hspace{-1mm} + \hspace{-1mm} \varphi_3 \hspace{-1mm} + 
 \hspace{-1mm} \delta \hspace{-1mm} + \hspace{-1mm} \varepsilon )^{L,R},  \hspace{2mm} 
\varphi^{L,R}_{32} = -(  \varphi_1 \hspace{-1mm} + \hspace{-1mm} \varphi_3 \hspace{-1mm}
 + \hspace{-1mm} \delta \hspace{-1mm} + \hspace{-1mm} \varepsilon )^{L,R},  \hspace{2mm} 
\varphi^{L,R}_{33} = -(  \varphi_1 \hspace{-1mm} + \hspace{-1mm} \varphi_2 \hspace{-1mm} 
 + \hspace{-1mm} \varepsilon )^{L,R}, 
\label{eq:phi3i}  
\end{eqnarray}
$s^{L,R}_{ij} = \sin \theta^{L,R}_{ij}, \, c^{L,R}_{ij} = \cos \theta^{L,R}_{ij}$, \,
$\theta^{L,R}_{12}, \, \theta^{L,R}_{23}, \, \theta^{L,R}_{13}$ and 
$\delta^{L,R}, \, \varepsilon^{L,R}, \, \varphi^{L,R}_0, \, \varphi^{L,R}_1, \, 
\varphi^{L,R}_2, \, \varphi^{L,R}_3 $ 
are the arbitrary angles and phases of the matrices $K^{L,R}_2$. 
Keeping in mind that the phases of the quark and lepton states are fixed by 
the standard choice of the matrices $V_{CKM}$ and $U_{PMNS}$ 
for the matrices $K^{L,R}_2$ we use the general 
form~(\ref{eq:KLR2})--(\ref{eq:phi3i}). 

With account of (\ref{eq:KLR2})--(\ref{eq:phi3i}) the factors 
$\beta_{P,i j }^2$, $|k^L_{P,33}-k^R_{P,33}|^2$ 
can be expressed in terms of mixing angles and phases of the matrices $K^{L,R}_2$. 
For $P = K^0_L$ the expressions $\beta_{K^0_L,i j}^2$ have the form  
\begin{eqnarray}
&&
\hspace{-15mm} 
\beta_{K^0_L,11}^2 
= \big( \, | (s^{L}_{12}c^{L}_{23}e^{i\varepsilon^{L}} + 
c^{L}_{12}s^{L}_{23}s^{L}_{13}e^{i(\delta^{L} + \varepsilon^{L})} ) \, 
c^{R}_{12}c^{R}_{13} + 
\nonumber 
\\
&&
+ \, (s^{R}_{12}c^{R}_{23}e^{-i\varepsilon^{R}} + 
c^{R}_{12}s^{R}_{23}s^{R}_{13}e^{-i(\delta^{R} + \varepsilon^{R})} ) \, 
c^{L}_{12}c^{L}_{13} \, |^2 + L \leftrightarrow R \big) / 4 ,
\label{eq:beta2K0L11}  
\\
&&
\hspace{-15mm} 
\beta_{K^0_L,22}^2 
= \big( \, | (c^{L}_{12}c^{L}_{23}e^{i\varepsilon^{L}} - 
s^{L}_{12}s^{L}_{23}s^{L}_{13}e^{i(\delta^{L} + \varepsilon^{L})} ) \, 
s^{R}_{12}c^{R}_{13} + 
\nonumber 
\\
&&
+ \, (c^{R}_{12}c^{R}_{23}e^{-i\varepsilon^{R}} - 
s^{R}_{12}s^{R}_{23}s^{R}_{13}e^{-i(\delta^{R} + \varepsilon^{R})} ) \, 
s^{L}_{12}c^{L}_{13} \, |^2 + L \leftrightarrow R \big) / 4 ,
\label{eq:beta2K0L22}  
\\
&&
\hspace{-15mm} 
\beta_{K^0_L,12}^2 = \beta_{K^0_L,21}^2  
= \big( \, | (s^{L}_{12}c^{L}_{23}e^{i\varepsilon^{L}} + 
c^{L}_{12}s^{L}_{23}s^{L}_{13}e^{i(\delta^{L} + \varepsilon^{L})} ) \, 
s^{R}_{12}c^{R}_{13} + 
\nonumber 
\\
&&
+ \, (-c^{R}_{12}c^{R}_{23}e^{-i\varepsilon^{R}} + 
s^{R}_{12}s^{R}_{23}s^{R}_{13}e^{-i(\delta^{R} + \varepsilon^{R})} ) \, 
c^{L}_{12}c^{L}_{13} \, |^2 + L \leftrightarrow R \big) / 4 .  
\label{eq:beta2K0L12}  
\end{eqnarray}                                                   

The mixing factors 
$\beta_{P,i j }^2$, $|k^L_{P,33}-k^R_{P,33}|^2$ 
for $P =(B^0, \, B_s )$ are more complicated and can be presented in the form 
\begin{eqnarray}
&&
\hspace{-15mm} 
\beta_{B^0,11}^2 \hspace{-1mm} = \hspace{-1mm} \big( \, | s^{L}_{12}s^{L}_{23} -  
c^{L}_{12}c^{L}_{23}s^{L}_{13}e^{i\delta^{L}} |^2  \, (c^{R}_{12}c^{R}_{13})^2 + 
L \leftrightarrow R \, \big) / 2, 
\label{eq:beta2B011}  
\\
&&
\hspace{-15mm} 
\beta_{B^0,22}^2 \hspace{-1mm}  = \hspace{-1mm}  \big( \, | c^{L}_{12}s^{L}_{23} +  
s^{L}_{12}c^{L}_{23}s^{L}_{13}e^{i\delta^{L}} |^2  \, (s^{R}_{12}c^{R}_{13})^2 + 
L \leftrightarrow R \, \big) / 2, 
\label{eq:beta2B022}  
\\
&&
\hspace{-15mm} 
\beta_{B^0,12}^2  \hspace{-1mm} = \hspace{-1mm}  \big( \, | s^{L}_{12}s^{L}_{23} -  
c^{L}_{12}c^{L}_{23}s^{L}_{13}e^{i\delta^{L}} |^2  \, (s^{R}_{12}c^{R}_{13})^2 + 
L \leftrightarrow R \, \big) / 2, 
\label{eq:beta2B012}  
\\
&&
\hspace{-15mm} 
\beta_{B^0,21}^2  \hspace{-1mm} = \hspace{-1mm}  \big( \, | c^{L}_{12}s^{L}_{23} +  
s^{L}_{12}c^{L}_{23}s^{L}_{13}e^{i\delta^{L}} |^2  \, (c^{R}_{12}c^{R}_{13})^2 + 
L \leftrightarrow R \, \big) / 2, 
\label{eq:beta2B021}  
\\
&&
\hspace{-15mm} 
\beta_{B^0,31}^2  \hspace{-1mm} = \hspace{-1mm}  \big( \, | c^{L}_{23}c^{L}_{13} - 
c^{R}_{23}c^{R}_{13} e^{i(\chi^{R}_1-\varepsilon^{R}-\chi^{L}_1+\varepsilon^{L})}
m_{\tau}/(2 \bbar{m}_{B^0} R_{B^0}^V) \, |^2  \, (c^{R}_{12}c^{R}_{13})^2 +    
L \leftrightarrow R \, \big) / 2, 
\label{eq:beta2B031}  
\\
&&
\hspace{-15mm} 
\beta_{B^0,32}^2  \hspace{-1mm} = \hspace{-1mm}  \big( \, | c^{L}_{23}c^{L}_{13} - 
c^{R}_{23}c^{R}_{13} e^{i(\chi^{R}_1-\varepsilon^{R}-\chi^{L}_1+\varepsilon^{L})}
m_{\tau}/(2 \bbar{m}_{B^0} R_{B^0}^V) \, |^2  \, (s^{R}_{12}c^{R}_{13})^2 +    
L \leftrightarrow R \, \big) / 2, 
\label{eq:beta2B032}  
\\
&&
\hspace{-15mm} 
\beta_{B^0,13}^2 \hspace{-1mm}  = \hspace{-1mm}  \big( \, |s^{R}_{13} - 
s^{L}_{13}e^{i(\chi^{R}_2-\chi^{L}_2)} m_{\tau}/(2 \bbar{m}_{B^0} R_{B^0}^V) \, |^2 \, 
| s^{L}_{12}s^{L}_{23} - c^{L}_{12}c^{L}_{23}s^{L}_{13}e^{i\delta^{L}} |^2  + 
L \leftrightarrow R \, \big) / 2, 
\label{eq:beta2B013}  
\\
&&
\hspace{-15mm} 
\beta_{B^0,23}^2  \hspace{-1mm} = \hspace{-1mm}  \big( \, |s^{R}_{13} - 
s^{L}_{13}e^{i(\chi^{R}_2-\chi^{L}_2)} m_{\tau}/(2 \bbar{m}_{B^0} R_{B^0}^V) \, |^2 \, 
| c^{L}_{12}s^{L}_{23} + s^{L}_{12}c^{L}_{23}s^{L}_{13}e^{i\delta^{L}} |^2  + 
L \leftrightarrow R \, \big) / 2, 
\label{eq:beta2B023}  
\\
&&
\hspace{-15mm} 
\beta_{B^0,33}^2 \hspace{-1mm} = \hspace{-1mm} 
\big( \, | c^{L}_{23}c^{L}_{13}s^{R}_{13} \hspace{-1mm}-\hspace{-1mm}  
[ \, c^{L}_{23}c^{L}_{13}s^{L}_{13} 
e^{i(\chi^{R}_2-\chi^{L}_2)} \hspace{-1mm}+\hspace{-1mm}  
c^{R}_{23}c^{R}_{13}s^{R}_{13} 
e^{i(\chi^{R}_1-\varepsilon^{R}-\chi^{L}_1+\varepsilon^{L})} \, ] 
m_{\tau}/(2 \bbar{m}_{B^0} R_{B^0}^V) \, |^2  \hspace{-1mm} + \hspace{-1mm}  
\nonumber 
\\
&&
+ L  \leftrightarrow  R \, \big) / 2, 
\label{eq:beta2B033}  
\\
&&
\hspace{-15mm} 
| k^{L}_{B^0,33} - k^{R}_{B^0,33}|^2 \hspace{-1mm} = \hspace{-1mm} 
| \, c^{L}_{23}c^{L}_{13}s^{R}_{13} -   
c^{R}_{23}c^{R}_{13}s^{L}_{13} 
e^{i(\chi^{R}_1+\chi^{R}_2-\varepsilon^{R}-\chi^{L}_1-\chi^{L}_2+\varepsilon^{L})} \,|^2,  
\label{eq:KLmKR2B0}  
\end{eqnarray}
and 
\begin{eqnarray}
&&
\hspace{-15mm} 
\beta_{B_s,11}^2 \hspace{-1mm} = \hspace{-1mm} \big( \, 
| s^{L}_{12}s^{L}_{23} - c^{L}_{12}c^{L}_{23}s^{L}_{13}e^{i\delta^{L}} |^2 \,
| s^{R}_{12}c^{R}_{23} + c^{R}_{12}s^{R}_{23}s^{R}_{13}e^{-i\delta^{R}} |^2 \, +
\, L \leftrightarrow R \, \big) / 2, 
\label{eq:beta2Bs11}  
\\
&&
\hspace{-15mm} 
\beta_{B_s,22}^2 \hspace{-1mm} = \hspace{-1mm} \big( \, 
| c^{L}_{12}s^{L}_{23} + s^{L}_{12}c^{L}_{23}s^{L}_{13}e^{i\delta^{L}} |^2 \,
| c^{R}_{12}c^{R}_{23} - s^{R}_{12}s^{R}_{23}s^{R}_{13}e^{-i\delta^{R}} |^2 \, +
\, L \leftrightarrow R \, \big) / 2, 
\label{eq:beta2Bs22}  
\\
&&
\hspace{-15mm} 
\beta_{B_s,12}^2 \hspace{-1mm} = \hspace{-1mm} \big( \, 
| s^{L}_{12}s^{L}_{23} - c^{L}_{12}c^{L}_{23}s^{L}_{13}e^{i\delta^{L}} |^2 \,
| c^{R}_{12}c^{R}_{23} - s^{R}_{12}s^{R}_{23}s^{R}_{13}e^{-i\delta^{R}} |^2 \, +
\, L \leftrightarrow R \, \big) / 2, 
\label{eq:beta2Bs12}  
\\
&&
\hspace{-15mm} 
\beta_{B_s,21}^2 \hspace{-1mm} = \hspace{-1mm} \big( \, 
| c^{L}_{12}s^{L}_{23} + s^{L}_{12}c^{L}_{23}s^{L}_{13}e^{i\delta^{L}} |^2 \,
| s^{R}_{12}c^{R}_{23} + c^{R}_{12}s^{R}_{23}s^{R}_{13}e^{-i\delta^{R}} |^2 \, +
\, L \leftrightarrow R \, \big) / 2, 
\label{eq:beta2Bs21}  
\\
&&
\hspace{-15mm} 
\beta_{B_s,31}^2  \hspace{-1mm} = \hspace{-1mm}  \big( \, 
| c^{L}_{23}c^{L}_{13} - 
c^{R}_{23}c^{R}_{13} e^{i(\chi^{R}_1-\varepsilon^{R}-\chi^{L}_1+\varepsilon^{L})}
m_{\tau}/(2 \bbar{m}_{B_s} R_{B_s}^V) |^2  \, 
| s^{R}_{12}c^{R}_{23} + c^{R}_{12}s^{R}_{23}s^{R}_{13}e^{-i\delta^{R}} |^2 + 
\nonumber 
\\
&&
+ \, L \leftrightarrow R \, \big) / 2, 
\label{eq:beta2Bs31}  
\\
&&
\hspace{-15mm} 
\beta_{B_s,32}^2  \hspace{-1mm} = \hspace{-1mm}  \big( \, 
| c^{L}_{23}c^{L}_{13} - 
c^{R}_{23}c^{R}_{13} e^{i(\chi^{R}_1-\varepsilon^{R}-\chi^{L}_1+\varepsilon^{L})}
m_{\tau}/(2 \bbar{m}_{B_s} R_{B_s}^V) |^2  \, 
| c^{R}_{12}c^{R}_{23} - s^{R}_{12}s^{R}_{23}s^{R}_{13}e^{-i\delta^{R}} |^2 + 
\nonumber 
\\
&&
+ \, L \leftrightarrow R \, \big) / 2, 
\label{eq:beta2Bs32}  
\\
&&
\hspace{-15mm} 
\beta_{B_s,13}^2  \hspace{-1mm} = \hspace{-1mm}  \big( \, 
| s^{R}_{23}c^{R}_{13} \hspace{-1mm} - \hspace{-1mm}  
s^{L}_{23}c^{L}_{13} e^{i(\chi^{R}_2+\delta^{R}+\varepsilon^{R}-
\chi^{L}_2-\delta^{L}-\varepsilon^{L})}
m_{\tau}/(2 \bbar{m}_{B_s} R_{B_s}^V) \, |^2  \, 
| s^{L}_{12}s^{L}_{23} \hspace{-1mm} - \hspace{-1mm} c^{L}_{12}c^{L}_{23}s^{L}_{13}
e^{i\delta^{L}} |^2 \hspace{-1mm} + 
\nonumber 
\\
&&
+ \, L \leftrightarrow R \, \big) / 2, 
\label{eq:beta2Bs13}  
\\
&&
\hspace{-15mm} 
\beta_{B_s,23}^2  \hspace{-1mm} = \hspace{-1mm}  \big( \, 
| s^{R}_{23}c^{R}_{13} \hspace{-1mm} - \hspace{-1mm} 
s^{L}_{23}c^{L}_{13} e^{i(\chi^{R}_2+\delta^{R}+\varepsilon^{R}-
\chi^{L}_2-\delta^{L}-\varepsilon^{L})}
m_{\tau}/(2 \bbar{m}_{B_s} R_{B_s}^V) \, |^2  \, 
| c^{L}_{12}s^{L}_{23} \hspace{-1mm} + \hspace{-1mm} s^{L}_{12}c^{L}_{23}s^{L}_{13}
e^{i\delta^{L}} |^2 \hspace{-1mm} + 
\nonumber 
\\
&&
+ \, L \leftrightarrow R \, \big) / 2, 
\label{eq:beta2Bs23}  
\\
&&
\hspace{-15mm} 
\beta_{B_s,33}^2 \hspace{-1mm} = \hspace{-1mm} 
\big( \, 
| \, c^{L}_{23}c^{L}_{13}s^{R}_{23}c^{R}_{13} -  
[ \, c^{L}_{23}(c^{L}_{13})^2s^{L}_{23} 
e^{i(\chi^{R}_2+\delta^{R}+\varepsilon^{R}-\chi^{L}_2-\delta^{L}-\varepsilon^{L})} + 
\nonumber 
\\
&&
 + \, c^{R}_{23}(c^{R}_{13})^2s^{R}_{23} 
e^{i(\chi^{R}_1-\varepsilon^{R}-\chi^{L}_1+\varepsilon^{L})} \, ] 
m_{\tau}/(2 \bbar{m}_{B_s} R_{B_s}^V) \, |^2   +   
\, L  \leftrightarrow  R \, \big) / 2, 
\label{eq:beta2Bs33}  
\\
&&
\hspace{-15mm} 
| k^{L}_{B_s,33} - k^{R}_{B_s,33}|^2  =  
| \, c^{L}_{23}c^{L}_{13}s^{R}_{23}c^{R}_{13}  -   
c^{R}_{23}c^{R}_{13}s^{L}_{23}c^{L}_{13} 
e^{i(\chi^{R}_1+\chi^{R}_2+\delta^{R}-\chi^{L}_1-\chi^{L}_2-\delta^{L})} \,|^2 ,    
\label{eq:KLmKR2Bs}  
\end{eqnarray}
where 
\begin{eqnarray}
&&
\chi^{L,R}_1 = \varphi^{L,R}_0 - \varphi^{L,R}_1 - \varphi^{L,R}_2, \hspace{8mm}
\chi^{L,R}_2 = \varphi^{L,R}_0 + \varphi^{L,R}_3. 
\label{eq:chiLR12}  
\end{eqnarray}

The mixing factors (\ref{eq:beta2Pll1a}), (\ref{eq:beta2Pll1b}), 
(\ref{eq:beta2K0L11})--(\ref{eq:KLmKR2Bs}) describe in the general form  
the effect of the fermion mixing in leptoquark currents 
in the case (\ref{eq:neglectmemmu}) on the branching 
ratios $Br_V(P~\to~l \, l^{\prime})$ of the decays of pseudoscalar 
mesons $P =( K^0_L, \, B^0, \, B_s )$ into lepton-antilepton pairs.  
These mixing factors can be used for the analysis of the branching 
ratios (\ref{eq:Br_VPll1f})--(\ref{eq:Br_VPtauptaumf}) in dependence on the mixing 
angles and phases of the mixing matrices 
(\ref{eq:KLR2}).

As seen from the (\ref{eq:beta2B031})--(\ref{eq:KLmKR2B0}), 
(\ref{eq:beta2Bs31})--(\ref{eq:KLmKR2Bs}) the mixing factors 
$\beta_{P,e \tau}^2$, \,  $\beta_{P,\mu \tau}^2$, \, 
$\beta_{P,\tau^+\tau^-}^2$, \, $| k^{L}_{P,33}~-~k^{R}_{P,33}|^2$ 
for $P =(B^0, \, B_s )$ depend on the phases 
$\varphi^{L,R}_0$,  $\varphi^{L,R}_1$,  $\varphi^{L,R}_2$,  $\varphi^{L,R}_3 $ 
through 
their combinations~(\ref{eq:chiLR12}).    
With fixed values of mixing angles 
$\theta^{L,R}_{12}, \, \theta^{L,R}_{23}, \, \theta^{L,R}_{13}$  
and phases~$\delta^{L,R}$ these mixing factors can be minimized over 
phases~(\ref{eq:chiLR12}) by the conditions          
\begin{eqnarray}
&& 
\chi^{L}_1 - \varepsilon^{L} - \chi^{R}_1 + \varepsilon^{R} = 0, \hspace{7mm}
\chi^{L}_2 - \chi^{R}_2 = 0,  \hspace{7mm}
\delta^{L} + \varepsilon^{L} - \delta^{R} -\varepsilon^{R} = 0.  
\label{eq:hi21edLR0}  
\end{eqnarray}

The most stringent lower mass limits for vector leptoquark are resulting 
from the experimental data on the branching ratios $Br_V(K^0_L \to l \, l^{\prime})$. 
As a result for the more small masses $m_V$ 
the mixing factors $\beta_{K^0_L,ll^{\prime}}^2$ must be very small (close to zero)      
and can be assumed in the further analysis to be equal to zero     
\begin{eqnarray}
&& 
\hspace{-15mm} 
\beta_{K^0_L,ll^{\prime}}^2 = 0 
\label{eq:betaK0Lll1to0}  
\end{eqnarray}
for $l \, l^{\prime} = e^+ e^-, \, \mu^+ \mu^-, \, e \mu$,  
where $\beta_{K^0_L,ll^{\prime}}^2$ 
are given by the relations (\ref{eq:beta2Pll1a}), (\ref{eq:beta2Pll1b}) and 
(\ref{eq:beta2K0L11})--(\ref{eq:beta2K0L12}).  

There are two solutions of the equations (\ref{eq:betaK0Lll1to0}): 
\begin{eqnarray}
&&\hspace{-15mm}
\mbox{solution A:} \hspace{10mm}
\theta^{L}_{23}=\theta^{R}_{23}= \pi /2, \hspace{5mm}
\theta^{L}_{13}=\theta^{R}_{13}\equiv\theta_{13}, \hspace{5mm}
\delta^{L} + \varepsilon^{L} + \delta^{R} +\varepsilon^{R} = \pi
\label{eq:solutionA}  
\end{eqnarray}
and 
\begin{eqnarray}
&&\hspace{-95mm}
\mbox{solution B:} \hspace{10mm}   
\theta^{L}_{13}=\theta^{R}_{13} = \pi /2.
\label{eq:solutionB}  
\end{eqnarray}

In both cases (\ref{eq:solutionA}) and (\ref{eq:solutionB}) the mixing factors  
$\beta_{P,33}^2$ and $| k^{L}_{P,33} - k^{R}_{P,33}|^2$ 
 for $P =(B^0, \, B_s )$ are equal to zero, which gives that in these cases  
$Br_V(B^0\hspace{-1mm}\to\hspace{-1mm}\tau^+\tau^-)\hspace{-1mm}=\hspace{-1mm}
Br_V(B_s\hspace{-1mm}\to\hspace{-1mm}\tau^+\tau^-)\hspace{-1mm}=0$.   

In the case (\ref{eq:hi21edLR0}), (\ref{eq:solutionA}) we obtain from 
(\ref{eq:beta2B011})--(\ref{eq:KLmKR2Bs})
the nonzero mixing factors (\ref{eq:beta2Pll1a}), (\ref{eq:beta2Pll1b}) in the form   
\begin{eqnarray}
&&
\beta_{B^0,e^+ e^-}^2 =  \beta_{B^0,\mu^+ \mu^-}^2 = 
c_{13}^2 \big( \, (s^{L}_{12} c^{R}_{12})^2 + (s^{R}_{12} c^{L}_{12})^2 \, \big) / 2, 
\label{eq:beta2B0epemmupmumMA}  
\\
&&
\beta_{B^0,e \mu}^2 =   
c_{13}^2 \big( \, (s^{L}_{12} s^{R}_{12})^2 + (c^{L}_{12} c^{R}_{12})^2 \, \big) , 
\label{eq:beta2B0emuMA}  
\\
&&
\beta_{B^0,e \tau}^2 = \big( \, 1 -m_{\tau}/(2 \bbar{m}_{B^0} R_{B^0}^V) \, \big)^2 \,   
s_{13}^2 \big( \, (s^{L}_{12})^2 + (s^{R}_{12})^2 \, \big) / 2, 
\label{eq:beta2B0etauMA}  
\\
&&
\beta_{B^0,\mu \tau}^2 = \big( \, 1 -m_{\tau}/(2 \bbar{m}_{B^0} R_{B^0}^V) \, \big)^2 \,   
s_{13}^2 \big( \, (c^{L}_{12})^2 + (c^{R}_{12})^2 \, \big) / 2, 
\label{eq:beta2B0mutauMA}  
\end{eqnarray}
and 
\begin{eqnarray}
&&
\beta_{B_s,e^+ e^-}^2 =  \beta_{B_s,\mu^+ \mu^-}^2 = 
s_{13}^2 \big( \, (s^{L}_{12} c^{R}_{12})^2 + (s^{R}_{12} c^{L}_{12})^2 \, \big) / 2, 
\label{eq:beta2BsepemmupmumMA}  
\\
&&
\beta_{B_s,e \mu}^2 =   
s_{13}^2 \big( \, (s^{L}_{12} s^{R}_{12})^2 + (c^{L}_{12} c^{R}_{12})^2 \, \big) , 
\label{eq:beta2BsemuMA}  
\\
&&
\beta_{B_s,e \tau}^2 = \big( \, 1 -m_{\tau}/(2 \bbar{m}_{B_s} R_{B_s}^V) \, \big)^2 \,   
c_{13}^2 \big( \, (s^{L}_{12})^2 + (s^{R}_{12})^2 \, \big) / 2, 
\label{eq:beta2BsetauMA}  
\\
&&
\beta_{B_s,\mu \tau}^2 = \big( \, 1 -m_{\tau}/(2 \bbar{m}_{B_s} R_{B_s}^V) \, \big)^2 \,   
c_{13}^2 \big( \, (c^{L}_{12})^2 + (c^{R}_{12})^2 \, \big) / 2. 
\label{eq:beta2BsmutauMA}  
\end{eqnarray}
As seen from (\ref{eq:beta2B0epemmupmumMA})--(\ref{eq:beta2BsmutauMA})  
the mixing factors $\beta_{B^0,ll^{\prime}}^2$, $\beta_{B_s,ll^{\prime}}^2$ 
in the case (\ref{eq:hi21edLR0}), (\ref{eq:solutionA}) depend in general case 
on three mixing angles $\theta_{13}, \,\, \theta_{12}^{L}, \,\, \theta_{12}^{R}$.   

In the case (\ref{eq:hi21edLR0}), (\ref{eq:solutionB}) the mixing factors 
(\ref{eq:beta2Pll1a}), (\ref{eq:beta2Pll1b}) 
depend on two effective mixing angles $\bar{\theta}^{L}, \, \bar{\theta}^{R} $ 
with $\sin \bar{\theta}^{L,R} =  | c_{23}c_{12} - s_{23}s_{12}e^{i\delta} |^{L,R}$   
and can be obtained from (\ref{eq:beta2B0epemmupmumMA})--(\ref{eq:beta2BsmutauMA}) 
by the substitutions 
$\theta_{13} \to \pi/2, \,\,\,   \theta_{12}^{L,R} \to \bar{\theta}^{L,R}$.   

The branching ratios (\ref{eq:Br_VPll1f})--(\ref{eq:Br_VPtauptaumf}) 
with the factors $\beta_{P,ll^{\prime}}^2$ depending on the mixing angles 
have been numerically analysed 
with account of experimental data 
on the decays (\ref{eq:KL0B0Bsdecays}). We vary the mixing angles 
$\theta_{13}, \, \theta^{L}_{12}, \, \theta^{R}_{12}$  
in the case (\ref{eq:hi21edLR0}), (\ref{eq:solutionA}) 
and the mixing angles $\bar{\theta}^{L}, \, \bar{\theta}^{R} $ 
in the case (\ref{eq:hi21edLR0}, (\ref{eq:solutionB})  
to find the minimal vector leptoquark mass $m_V$ satisfying these data. 

In the numerical analysis we use the data on the masses   
$m_{l_{i}}, \,\, m_{d_{i}}$ of leptons and quarks, 
the data on the masses $m_{P}$, 
life times $\tau_{P}$ ($\tau_{P} \rightarrow \Gamma_P^{tot} $)    
and the form factors $f_P$   
\begin{eqnarray}
&& 
f_{K^0_L} = f_{K^-} = 155.72 \, MeV, \,\,
f_{B^0} = 190.9 \, MeV, \,\, f_{B_s}= 227.2\, MeV
\nonumber 
\label{eq:fPexp}  
\end{eqnarray}
of mesons $P =( K^0_L, \, B^0, \, B_s ) $
from ref. \cite{RPP2016}. 
The experimental data on the branching ratios   
$Br(P \to l \, l^{\prime})^{exp}$ are also taken from the ref. \cite{RPP2016}  
except the branching ratios of the decays 
$ B^0, B_s  \to \mu^+\mu^-, \tau^+\tau^- $ for which we use 
the current data 
\begin{eqnarray}
&&  
Br(B^0 \to \mu^+ \mu^-)^{exp} < 3.4\cdot 10^{-10}  
\mbox{\hspace{2mm} \cite{Aaij:2017vad}}, 
\label{eq:BrB0mupmumexp}  
\\
&&  
Br(B^0 \to \tau^+\tau^-)^{exp} < 2.1\cdot 10^{-3}  
\mbox{\hspace{2mm} \cite{Aaij:2017xqt}}, 
\label{eq:BrB0tauptaumexp}  
\\
&&  
Br(B_s \to \mu^+ \mu^-)^{exp} = (3.0 \pm 0.6^{+0.3}_{-0.2})\cdot 10^{-9} 
\mbox{\hspace{2mm} \cite{Aaij:2017vad}}, 
\label{eq:BrBsmupmumexp}  
\\
&&  
Br(B_s \to \tau^+\tau^-)^{exp} < 6.8\cdot 10^{-3} 
\mbox{\hspace{2mm} \cite{Aaij:2017xqt}} 
\label{eq:BrBstauptaumexp}  
\end{eqnarray}
of refs.\cite{Aaij:2017vad,Aaij:2017xqt} 
and the branching ratios of the decays 
$ B^0, B_s  \to  e \mu $ for which we use the recent data
\begin{eqnarray}
&&  
Br(B^0 \to e\mu)^{exp} < 1.0\cdot 10^{-9}  
\mbox{\hspace{2mm} \cite{Aaij:2017newdata}}, 
\label{eq:BrB0emuexp}  
\\
&&  
Br(B_s \to e\mu)^{exp} < 5.4\cdot 10^{-9}  
\mbox{\hspace{2mm} \cite{Aaij:2017newdata}} 
\label{eq:BrBsemuexp}  
\end{eqnarray}
of the ref.\cite{Aaij:2017newdata}. 
For the mass scale $M_c$ of the $SU_V(4)$ symmetry breaking  
we choose the value $M_c = 100 \, TeV$.     

Using  (\ref{eq:beta2B0epemmupmumMA})--(\ref{eq:beta2BsmutauMA})
for the case (\ref{eq:solutionA}) we have found the lower vector leptoquark mass limit 
\begin{eqnarray}
&& 
 m_V > 86 \,\, TeV  
\label{eq:mVlim}  
\end{eqnarray}
in the case of the chiral mixing and the mass limit $ m_V > 87.4 \,\, TeV $ in 
the case of the vectorkike ($\theta_{12}^{L}=\theta_{12}^{R} \equiv \theta_{12} $) one. 
The case (\ref{eq:solutionB}) with the chiral and 
vectorkike ($\bar{\theta}^{L}=\bar{\theta}^{R}$) mixings 
gives the mass limits $ m_V > 89.3 \,\, TeV $ and  
$ m_V > 90.3 \,\, TeV $ respectively.  
As seen, the mass limit (\ref{eq:mVlim}) is the lowerest one.  

The effect of the fermion mixing on the decays of type (\ref{eq:KL0B0Bsdecays}) 
has been also considered in~\cite{KMS-JMPA2012} but for the particular choice of 
the mixing matrices correponding to the case~(\ref{eq:solutionA}) 
with the additional restrictions $\theta_{12}^{L}=\theta_{12}^{R}=\pi/2$. 
Besides, in comparing the theoretical results for $ B^0, B_s $ mesons 
with the experimental data only the first terms in the branching 
ratios (\ref{eq:Br_VPemu})--(\ref{eq:Br_VPmutau}) and in the mixing 
factors~(\ref{eq:beta2Pll1b}) have been taken into account, 
which contradicts the usual treatment of the experimental branching ratios 
as the sums over charge conjugated final states. 
By these two reasons the lower vector leptoquark mass 
limit $m_V > 38 \,\, TeV$ of ref.~\cite{KMS-JMPA2012} seems questionable 
(for example instead of this value the case~(\ref{eq:solutionA}) 
with $\theta_{12}^{L}=\theta_{12}^{R}=\pi/2$ 
gives in fact the mass limit $m_V > 101.3 \,\, TeV$).           

The mass limit (\ref{eq:mVlim}) is resulting from the current experimental data 
on the branching ratios $Br(P \to l \, l^{\prime})^{exp}$ of the ref. \cite{RPP2016}  
and (\ref{eq:BrB0mupmumexp})--(\ref{eq:BrBsemuexp})  
with account of the fermion mixing in leptoquark currents of the general form.   
As seen, it is essentially lower than that  
of order of $2\,000 \,\, TeV$ obtained with neglect of the fermion 
mixing  \cite{VW, KM1,KM2,AD-Mpla2007,AD-YaF2008} and noticeably exceeds 
the mass limit of order of $1 \,\, TeV$  
resulting from the current direct searches for the vector leptoquarks.      
It is worthy to note that the mass limit (\ref{eq:mVlim}) can be 
further lowered by account of the possible destructive interference of 
the contributions from the vector leptoquarks with those from the scalar 
leptoquarks which are also predicted in the scalar sector of the models  
with the four color symmetry. The possibility of such interference in 
the minimal model with the four color symmetry  
based on the gauge group $G_{MQLS}=SU_V(4) \times SU_L(2) \times U_R(1)$     
\cite{AD1, AD2,PW})
in the case of neglect of the fermion mixing ($K^L_2 = K^R_2 = I$)
has been demonstrated in~\cite{AD-Mpla2007,AD-YaF2008}.  
                                               
We have calculated the branching ratios $Br_V(P \to l \, l^{\prime})$ predicted 
by the vector leptoquarks with the lower allowed  mass $m_V = 86 \,\, TeV$.    
This value of the vector leptoquark mass is ensured by the appropriate values 
of the mixing angles $\theta_{13}, \, \theta^{L}_{12}, \, \theta^{R}_{12}$ from 
the region    
\begin{eqnarray}
&& 
\theta_{13} = 1.183 - 1.187,    \hspace{5mm}
\theta^{L}_{12} (\theta^{R}_{12}) = 0.00 (0.81) - 0.763 (\pi/2) 
\label{eq:theta_region}  
\end{eqnarray}
with the branching ratios $Br_V(P \to l \, l^{\prime})$ being invariant under 
exchange $\theta^{L}_{12} \leftrightarrow \theta^{R}_{12}$ 
in accordance with (\ref{eq:beta2B0epemmupmumMA})--(\ref{eq:beta2BsmutauMA}).      
The analysis shows that the variations of 
the branching ratios 
under variations 
of the mixing angles $\theta_{13}, \, \theta^{L}_{12}, \, \theta^{R}_{12}$ 
within the region~(\ref{eq:theta_region}) for the decays 
$ B^0, \, B_s \to e^+e^-, \mu^+ \mu^-, e \mu $ 
do not exceed 3\% and the branching ratios 
of the decays 
$ B^0 \to e \tau, \mu \tau $ and $ B_s \to e \tau, \mu \tau $ 
 are of order of $10^{-9}$ and  $10^{-10} - 10^{-9}$ respectively.      
For definiteness in the second column of the Table~\ref{tab:BrVPl1l2_newdata2017} 
we present the branching ratios $Br_V(P \to l \, l^{\prime})$ 
for $m_V = 86 \,\, TeV$ 
at $ \theta_{13} = 1.183, \, \theta^{L}_{12} (\theta^{R}_{12}) = 0.00 (0.81) $.     
In the third column of the Table~\ref{tab:BrVPl1l2_newdata2017} we present 
for comparision the current experimental data on the decays under consideration.   
\vspace{-5mm}
\begin{table}[h]
\caption{ Branching ratios $Br_V(P \to l \, l^{\prime})$ predicted 
by the vector leptoquarks with the lower mass $m_V = 86 \,\, TeV$   
with account of fermion mixing 
at $ (\theta_{13}) \theta^{L}_{12} (\theta^{R}_{12})=(1.183) 0.00 (0.81) $   
}
\vspace*{3mm}
 \centerline{
\epsfxsize=0.8 
\textwidth \epsffile{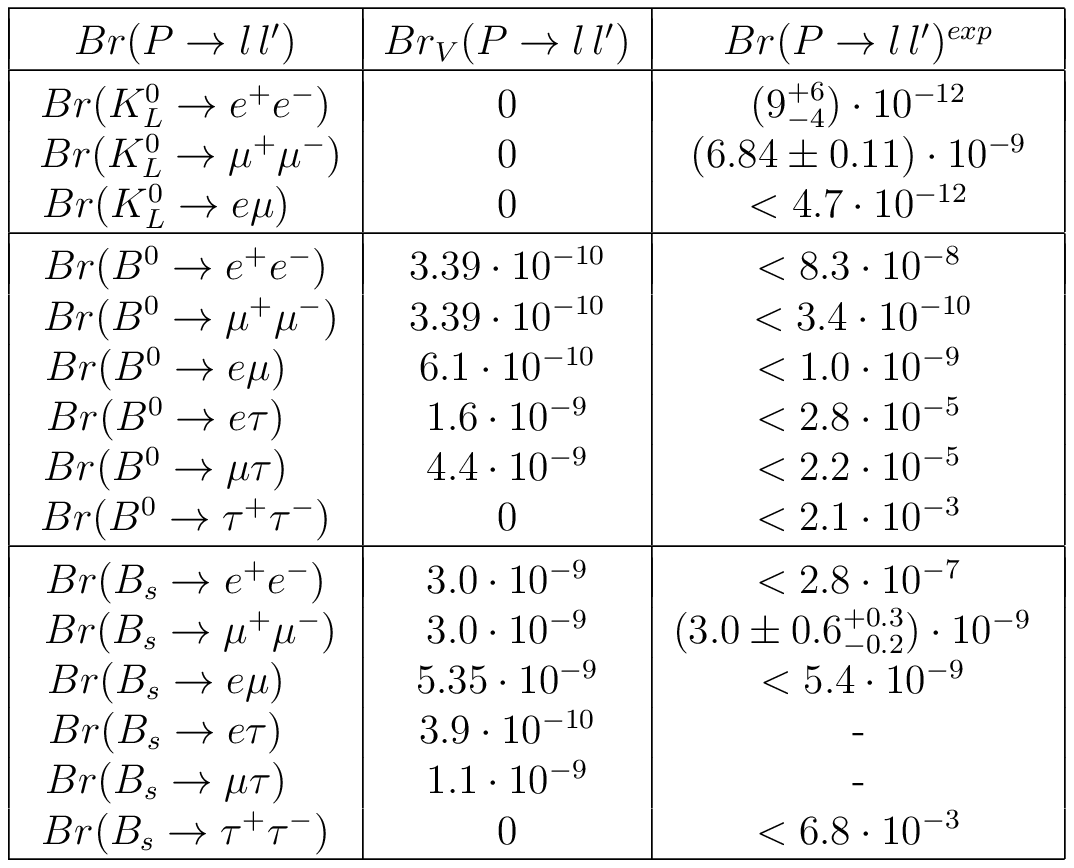}} 
\label{tab:BrVPl1l2_newdata2017}
\end{table}

As seen from the Table~\ref{tab:BrVPl1l2_newdata2017} the branching ratios  
predicted by the vector leptoquarks with $m_V = 86 \,\, TeV$ for the decays  
$ B^0, B_s \to \mu^+ \mu^-, e \mu $  are close to (or compatible with) 
the corresponding experimental data. It means that 
these decays are now most suitable for search for the vector leptoquarks 
and for setting the new more stringent limits on their mass.  
The decays $ B^0, B_s \to \mu^+ \mu^-, e \mu $ as the perspective ones for obtaining 
the new mass limit for the vector leptoquarks have been pointed out 
in ref.~\cite{AD-ProcQFTHEP2017} at the data of ref. \cite{RPP2016}  
and (\ref{eq:BrB0mupmumexp})--(\ref{eq:BrBstauptaumexp}),    
this conclusion is now confirmed at 
the data (\ref{eq:BrB0emuexp})--(\ref{eq:BrBsemuexp}) with improving 
the vector leptoquark mass limit from $ m_V > 78 \,\, TeV $ 
of ref.~\cite{AD-ProcQFTHEP2017} to the current mass limit~(\ref{eq:mVlim}).              

As concerns the decays  
$B^0 \to e^+ e^-, e \tau, \mu \tau$, and  $B_s \to e^+ e^-$ the current experimental 
limits on these decays essentially (by order of $2\div4$) exceed the corresponding 
expected contributions from the vector leptoquarks and the search for these decays 
will need the correspondinly high statistics.     

The experimental data on the decays $B_s \to  e \tau, \mu \tau$ by now are absent  
and the vector leptoquarks with mass $m_V = 86 \,\, TeV$  predict for these decays 
the branching ratios of order of $10^{-10}$ and $10^{-9}$ respectively.   

The contributions of the vector leptoquarks to branching ratios of the decays  
$B^0, B_s  \to \tau^+\tau^-$ are very small 
(under assumptions (\ref{eq:betaK0Lll1to0}) they are equal to zero) and 
the search for these decays seems to be difficult.     

The numerical predictions of the branching ratios of the decays  
$B^0, B_s \to l \, l^{\prime}$ presented in the Table~\ref{tab:BrVPl1l2_newdata2017} 
will be usefull in the current and future experimental 
searches for these decays. 

In conclusion we resume the results of the work. 

The contributions of the vector leptoquarks of Pati-Salam type to the branching 
ratios of $ K_L^0, B^0, B_s \to l \, l^{\prime} $ decays are calculated and analysed  
with account of the fermion mixing in the leptoguark currents of the general type. 
The general parametrizations of the mixing matrices in the leptoguark currents 
are proposed and the 
mixing factors in the branching ratios as the general functions on mixing 
angles and phases are found. 

With account of the fermion mixing in leptoquark currents of the general form 
the lower vector leptoquark mass limit $ m_V > 86 \,\, TeV $ is found 
from the current experimental data on the branching ratios of these decays.  

The numerical predictions of the branching ratios of the decays  
$B^0, B_s \to l \, l^{\prime}$ 
at the lower vector leptoquark mass $m_V = 86 \,\, TeV$ 
are obtained in comparison with the current experimental data.  
It is shown that for the decays $ B^0, B_s \to \mu^+ \mu^-, e \mu $  
the predicted branching ratios are close to 
the experimental data and these decays are most promising for the search 
for the vector leptoquarks and for setting 
the new more stringent vector leptoquark mass limits  
whereas the analogous predictions for the decays  
$B^0 \to e^+ e^-, e \tau, \mu \tau$, and  $B_s \to e^+ e^-$ 
are essentially (by order of $2\div4$) less   
than their current experimental limits and the search for these decays 
will need the more high statistics.     

For the decays $B_s \to  e \tau, \mu \tau$ 
the experimental data by now are absent,   
the expected branching ratios for these decays can be of order or less 
than $10^{-10}$ and $10^{-9}$ respectively.  

The numerical predictions of the branching ratios of the decays  
$B^0, B_s \to l \, l^{\prime}$ at the current lower vector leptoquark 
mass $m_V = 86 \,\, TeV$ will be usefull in the current and future experimental 
searches for these decays.



\vspace{3mm} {\bf Acknowledgment}

The paper is done within 
 ``YSU Initiative Scientific Researches'' (Project No. AAAA-A16-116070610023-3)

\vspace{3mm}



\begin{thebibliography}{99}

%
%

%
\bibitem{PS}
    J.~C.~Pati, A.~Salam, Phys.~Rev. \textbf{D10},~275~(1974). 

\bibitem{FM}
    H.~Fritzsch, P.~Minkowski, Ann.~Phys. \textbf{93},~193~(1975). 

\bibitem{CEG}
    M.~S.~Chanowitz, J.~Ellis, M.~K.~Gaillard, 
    Nucl.~Phys. \textbf{B128},~506~(1977). 

\bibitem{GN}
    H.~Georgi,~D.~V.~Nanopoulos, Nucl.~Phys. \textbf{B155},~52~(1979).   

\bibitem{GRS}
    F.~Gursey,~P.~Ramond,~P.~Sikivie, Phys.~Lett. \textbf{B60},~177~(1976).

\bibitem{BN}
    R.~Barbieri,~D.~V.~Nanopoulos, Phys.~Lett. \textbf{B91},~369~(1980). 

\bibitem{DFGKK}
    I.~Dorsner,~S.~Fajfer,~A.~Greljo,~J.~F.~Kamenik,~N.Kosnik, 
    Phys.~Rep., \textbf{641},~1~(2016).        

\bibitem{SMG}
    S.~Sahoo,~R.~Mohanta,~ A.~K.~Giri, Phys.~Rev. \textbf{D95},~035027~(2017).

\bibitem{AFG}
    N~ Assad,~B.~Fornal,~B.~Grinstein, Phys. Lett. B 777, 324 (2018). 

\bibitem{LGN}
    L.~D.~Luzio,~A.~Greljo,~M.~Nardecchia, 
    Phys.~Rev. \textbf{D96},~115011~(2017). 

\bibitem{CCL}
    L.~Calibbi,~A.~Crivellin,~T.~Li, Preprint PSI-PR-12-17, 
    arXiv:1709.00692 [hep-ph].  

\bibitem{BCFI}
     M.~Bordone,~C.~Cornella,~J.~Fuentes-Martin,~G.~Isidori, 
     Preprint ZU-TH-36/17,  
     arXiv:1712.01368 [hep-ph].

\bibitem{BMS}
     R.~Barbieri,~C.~W.~Murphy,~F.~Senia, 
     Eur.~Phys.~J. \textbf{C77},~N.1,~8~(2017).

\bibitem{BT}
     R.~Barbieri,~A.~Tesi, arXiv:1712.06844 [hep-ph]. 

\bibitem{VW}
    G.~Valencia, S.~Willenbrock, Phys.~Rev. \textbf{D50},~6843~(1994).

\bibitem{KM1}
    A.~V.~Kuznetsov, N.~V.~Mikheev, Phys.~Lett. \textbf{B329},~295~(1994).

\bibitem{KM2}
    A.~V.~Kuznetsov, N.~V.~Mikheev, 
        Phys. At. Nucl. \textbf{58},~2228~(1995). 

\bibitem{AD-Mpla2007}
     A.~D.~Smirnov, Mod.~Phys.~Lett. \textbf{A 22},~2353~(2007). 

\bibitem{AD-YaF2008}
         A.~D.~Smirnov, 
        Phys. At. Nucl. \textbf{71},~1470~(2008).


\bibitem{RPP2016}
    C. Patrignani et al. (Particle Data Group), 
    Chin. Phys. \textbf{C 40},~100001~(2016).

\bibitem{Aaij:2017vad}
    R. Aaij et al. (LHCb Collaboration), Phys. Rev. Lett. \textbf{118},~191801~(2017).

\bibitem{Aaij:2017xqt}
    R. Aaij et al. (LHCb Collaboration), Phys. Rev. Lett. \textbf{118},~251802~(2017). 

\bibitem{Aaij:2017newdata}
    R. Aaij et al. (LHCb Collaboration),  
    arXiv: 1710.04111[hep-ex].  

\bibitem{KMS-JMPA2012}
     A.~V.~Kuznetsov, N.~V.~Mikheev, A.~V.~Serghienko,   
     Int. J. Mod. Phys. \textbf{A 27},~1250062~(2012). 

\bibitem{AD1}
    A.~D.~Smirnov, Phys.~Lett. \textbf{B346},~297~(1995).

\bibitem{AD2}
        A.~D.~Smirnov, 
        Phys. At. Nucl.  \textbf{58},~2137~(1995). 
%


\bibitem{PW}
   P.~F.~Perez,~M.~B.~Wise, 
   Phys.~Rev. \textbf{D88},~057703~(2013). 


\bibitem{AD-ProcQFTHEP2017}
     A.~D.~Smirnov, 
    EPJ Web of Conf., \textbf{158},~02004~(2017).

\end{thebibliography}
\end{document}